\documentclass[lettersize,journal]{IEEEtran}
\usepackage{stmaryrd}
\usepackage{lipsum}  

\usepackage{amssymb,amsmath,amsfonts,wasysym,latexsym}
\usepackage[acronyms,shortcuts]{glossaries}
\usepackage{bm}
\usepackage{tikz}
\usetikzlibrary{patterns,arrows,decorations.pathreplacing}
\usepackage{psfrag}
\usepackage{amssymb}
\newcommand{\argmin}{\mathop{\rm argmin}}
\newcommand{\normof}[2]{\left\|#1\right\|_{#2}}
\newcommand{\fronorm}[1]{\normof{#1}{\rm F}}            
\newcommand{\blockron}{\hspace{0.05cm}|\hspace{-0.1cm}\otimes\hspace{-0.1cm}| \hspace{0.05cm}}
\usepackage{epstopdf}
\usepackage{amsmath}
\usepackage{scalefnt}
\usepackage{color}
\usepackage{xcolor}
\usepackage{algorithm}
\usepackage{algpseudocode}
\usepackage{cite}
\usepackage{caption}
\usepackage{subcaption}
\usepackage{caption}
\usepackage{multirow}
\usepackage{graphicx}

\definecolor{forestgreen}{rgb}{0, 0.5,0.5} 
\definecolor{darkgreen}{rgb}{0,0.392157,0} 
\newtheorem{remark}{\textit{Remark}}

\newcommand{\nmode}[2]{\big[\ma{\mathcal{#1}}\big]_{\left(#2\right)}}
\newcommand{\sumx}[2]{\sum\limits_{#1}^{#2}}
\newcommand{\bb}[1]{\mathbb{#1}}
\newcommand{\ten}[1]{\boldsymbol{\mathcal #1}}
\newcommand{\ma}[1]{\boldsymbol{#1}}
\definecolor{green}{rgb}{0.1,0.75,0.2}

\newacronym{2G}{2G}{second generation}
\newacronym{3G}{3G}{third generation}
\newacronym{4G}{4G}{fourth generation}
\newacronym{5G}{5G}{fifth generation}
\newacronym{B5G}{B5G}{beyond fifth generation}
\newacronym{6G}{6G}{sixth generation}
\newacronym{3GPP}{3GPP}{3$\text{rd}$~Generation Partnership Project}
\newacronym{LTE}{LTE}{long term evolution}
\newacronym{NR}{NR}{new radio}
\newacronym{LS}{LS}{least squares}

\newacronym{IRS}{IRS}{intelligent reconfigurable surface}
\newacronym{RIS}{RIS}{reconfigurable intelligent surface}
\newacronym{LIS}{LIS}{large intelligent surface}
\newacronym{SDS}{SDS}{software-defined surface}

\newacronym{D2D}{D2D}{device-to-device}
\newacronym{BS}{BS}{base station}
\newacronym{UE}{UE}{user equipment}

\newacronym{SU}{SU}{single-user}
\newacronym{MU}{MU}{multi-user}
\newacronym{SISO}{SISO}{single-input single-output}
\newacronym{MISO}{MISO}{multiple-input single-output}
\newacronym{SIMO}{SIMO}{single-input multiple-output}
\newacronym{MIMO}{MIMO}{multiple-input multiple-output}

\newacronym{CSI}{CSI}{channel state information}
\newacronym{LOS}{LOS}{line-of-sight}
\newacronym{NLOS}{NLOS}{non-line-of-sight}

\newacronym{QoS}{QoS}{quality-of-service}
\newacronym{SE}{SE}{spectral efficiency}
\newacronym{EE}{EE}{energy efficiency}
\newacronym{SINR}{SINR}{signal to interference plus noise ratio}
\newacronym{SNR}{SNR}{signal to noise ratio}

\newacronym{ProSe}{ProSe}{proximity services}
\newacronym{NSPS}{NSPS}{national security and public safety}

\newacronym{RRM}{RRM}{radio resource management}
\newacronym{MS}{MS}{mode selection}
\newacronym{RA}{RA}{resource allocation}
\newacronym{PC}{PC}{power control}

\newacronym{BCD}{BCD}{block coordinate descent}

\newacronym{RF}{RF}{radio frequency}
\newacronym{AWGN}{AWGN}{additive white Gaussian noise}
\newacronym{MRC}{MRC}{maximum ratio combining}

\newacronym{AF}{AF}{amplify-and-forward}
\newacronym{DF}{DF}{decode-and-forward}
\newacronym{DFT}{DFT}{discrete Fourier transform}
\newacronym{TX}{TX}{transmitter}
\newacronym{RX}{RX}{receiver}
\newacronym{ALS}{ALS}{alternating least squares}
\newacronym{BALS}{BALS}{bilinear alternating least squares}
\newacronym{SVD}{SVD}{singular value decomposition}
\newacronym{HOSVD}{HOSVD}{high order singular value decomposition}
\newacronym{THOSVD}{THOSVD}{truncated high order singular value decomposition}
\newacronym{PARAFAC}{PARAFAC}{PARAllel FACtors}
\newacronym{AOD}{AOD}{angle of departure}
\newacronym{AOA}{AOA}{angle of arrival}
\newacronym{URA}{URA}{uniform rectangular array} 
\newacronym{ADR}{ADR}{achievable data rate}
\newacronym{NMSE}{NMSE}{normalized mean square error}
\newacronym{SER}{SER}{symbol error rate}
\newacronym{LRA}{LRA}{low-rank approximation}

\newacronym{ULA}{ULA}{uniform linear array}
\newacronym{mmWave}{mmWave}{milimiter-wave}
\newacronym{CS}{CS}{compressed sensing}
\newacronym{OFDM}{OFDM}{orthogonal frequency division multiplexing}
\newacronym{PIN}{PIN}{positive-intrinsic-negative}
\newacronym{BD-RIS}{BD-RIS}{beyond diagonal reconfigurable intelligent surface}
\newacronym{LS-Kron}{LS-Kron}{least squares Kronecker factorization}

\newacronym{BTALS}{BTALS}{block Tucker alternating least squares}
\newacronym{BTKF}{BTKF}{block Tucker Kronecker factorization}
\newacronym{PALS}{PALS}{PARAFAC alternating least squares}
\newacronym{PKF}{PKF}{PARAFAC Khatri-Rao factorization}

\usepackage[none]{hyphenat}
\begin{document}
\title{ 
PARAFAC-Based Channel Estimation for Beyond Diagonal Reconfigurable Surfaces
} 

\author{Gilderlan Tavares de Araújo, Bruno Sokal, André L. F. de Almeida 
        
     \thanks{The authors are with Laboratory of Applied Signal Processing (LASP), Department of Teleinformatics Engineering, Federal University of Cear\'{a}, Fortaleza-CE, Brazil. E-mails: \{gilderlan,brunosokal,andre\}@gtel.ufc.br. This work was supported in part by the Brazil's National Council for Scientific and Technological Development (CNPq) (projects 406517/2022-3 and 303356/2025-1), and FUNCAP (project ITR-0214-00041.01.00/23).} 
 
}
\renewcommand{\arraystretch}{1.3} 

\maketitle


\begin{abstract}
Channel estimation is a central bottleneck in BD-RIS-assisted MIMO systems. The richer inter-element coupling that enables large performance gains also makes training and hardware control substantially harder than in diagonal RIS architectures. Existing estimators either target only cascaded channels or require block-by-block reconfiguration of the BD-RIS interconnections, which is costly and difficult to implement in practice. To overcome this limitation, we propose a pilot-assisted tensor framework for group-connected BD-RIS under a two-timescale protocol, where the scattering structure is designed as a low-rank PARAFAC model with fixed factor matrices. This design keeps the interconnection topology constant across blocks and updates only phase shifts, enabling practical operation without sacrificing estimation quality. Building on this structure, we develop a PARAFAC-based alternating least-squares (PALS) receiver that recovers the individual channels. Numerical results confirm that PALS delivers markedly lower composite-channel NMSE than conventional LS, matches the accuracy of state-of-the-art tensor receivers, and sharply reduces BD-RIS design complexity.
\end{abstract}


\section{Introduction}\label{Sec:Introduction}
Reconfigurable intelligent surfaces (RISs) are a key enabler for future wireless networks because they provide software-controlled manipulation of the propagation environment \cite{Di_Renzo2020,Rui_Zhang_2021}. In RIS-assisted MIMO systems, however, channel estimation remains a critical bottleneck, as the passive nature of RIS elements and the large dimensionality of cascaded links make pilot design and parameter recovery particularly challenging \cite{Lee_2022,Paulo_2023}. A major advance in this direction was reported in \cite{Araujo_SAM20}, \cite{gil2021}, which first established a formal connection between tensor decompositions and RIS channel estimation. Specifically, \cite{gil2021} showed that the received pilot measurements admit a structured multilinear model, enabling identifiability analysis and efficient tensor-based estimators to recover channel factors with reduced training overhead.

This tensor-estimation perspective was subsequently extended to BD-RIS systems in \cite{ClerckX_TSP_CE}, where the richer scattering architecture (group-connected/fully connected) was incorporated into the estimation model. Therein, the authors demonstrated that tensor methods can estimate individual channel components beyond the cascaded representation, but it also highlighted a practical limitation: competitive performance often depends on repeatedly redesigning or reconfiguring the BD-RIS scattering pattern across training blocks \cite{CE_BD_RIS,Li_TSP2025,WangTSP2025}. Related semi-blind tensor methods \cite{gil-asilomar,gil_OJSP} improve spectral efficiency, yet the hardware reconfiguration burden remains largely unresolved.
Therefore, a key open issue is to jointly achieve: i) individual channel estimation accuracy; ii) low training overhead; and iii) hardware-feasible BD-RIS operation with limited reconfiguration.

Motivated by this gap, this paper proposes a pilot-assisted tensor framework for channel estimation in group-connected BD-RIS systems under a two-timescale protocol. Our core idea is to model the BD-RIS training structure as a PARAFAC tensor \cite{Harshman} with fixed low-rank factors, so that the interconnection topology remains unchanged across blocks and only phase coefficients are updated. Compared with prior tensor-based designs that reconfigure the full scattering structure at each block, the proposed strategy is substantially more implementation-friendly. On top of this design, we derive a PARAFAC-based alternating least-squares receiver (ALS) that estimates the individual channels. Beyond its conceptual simplicity, PALS also leverages the well-known uniqueness properties of PARAFAC, which yield necessary and sufficient identifiability conditions for the involved channels---conditions not established in earlier works. Numerical results demonstrate that the proposed method preserves the estimation performance of state-of-the-art receivers while markedly reducing BD-RIS design complexity and outperforming conventional LS estimation.\footnote{\textit{Notation}: Scalars are denoted by lowercase letters ($a$), vectors by bold lowercase letters ($\mathbf{a})$, matrices by bold capital letters ($\mathbf{A})$, and tensors by calligraphic letters $(\mathcal{A})$. The transpose and Hermitian transpose of a matrix $\mathbf{A}$ are denoted by $\mathbf{A}^{\textrm{T}}$ and $\mathbf{A}^{\textrm{H}}$, respectively. $D_i(\mathbf{A})$ is a diagonal matrix holding the $i$-th row of $\mathbf{A}$ on its main diagonal. $\| \cdot \|_{\text{F}}$ is the Frobenius norm. The operator $\textrm{diag}(\mathbf{a})$ forms a diagonal matrix from its vector argument, while $\diamond$  denotes the Khatri-Rao product. $\mathbf{A}_{i.}$ denotes the $i$-th row of $\mathbf{A}$, and $\mathbf{I}_{J}$ is an identity matrix of size $J \times J$. 
}


\section{System Model}\label{Sec:System_Model}
Consider a \ac{MIMO} system assisted by a \ac{BD-RIS}, where the transmitter and receiver are equipped with $M_T$ and $M_R$ antennas, respectively, and the BD-RIS has $N$ elements. For simplicity, we assume that the direct transmitter--receiver link is blocked. We adopt the two-timescale protocol in \cite{gil2021}, \cite{ClerckX_TSP_CE}, in which the transmission interval is divided into $K$ consecutive blocks, each composed of $T$ time slots. The pilot matrix is repeated within each block, whereas the BD-RIS response is kept fixed during a block and updated only across blocks.
We assume a group-connected \ac{BD-RIS} architecture \cite{CE_BD_RIS,CLX_BD_GC_2023,CLX_MS_2023}, where the $N$ reflecting elements are partitioned into $Q$ groups, each containing $\bar{N}$ interconnected elements \cite{CE_BD_RIS}, such that $N = \bar{N}Q$. Under this structure, the scattering matrix at block $k$ is written as $\ma{S}_k = \text{bdiag}(\ma{S}^{(1)}_{k}, \ldots, \ma{S}^{(Q)}_{k}) \in \bb{C}^{N \times N}$. The pilot signal matrix received at block $k$ is then given by
\begin{equation}
    \label{eq:rec_sig_k_sum} \ma{Y}_k = \sumx{q=1}{Q} \ma{G}^{(q)}\ma{S}^{(q)}_{k}\ma{H}^{(q)\text{T}}\ma{X}^{\text{T}} + \ma{B}_k \in \bb{C}^{M_R \times T},
\end{equation}
where $\ma{H}^{(q)} \in \bb{C}^{M_T \times \bar{N}}$ and $\ma{G}^{(q)}\in \bb{C}^{M_R \times \bar{N}}$ are the involved channels associated with the $q$ BD-RIS group, respectively, that correspond to the block $q$-th of $\ma{H}\in \bb{C}^{M_T \times \bar{N}Q}$ and  $\ma{G} \in \bb{C}^{M_R \times \bar{N}Q}$, respectively, defined as follows:
\begin{align}
    \ma{H}^{(q)} &= \ma{H}_{.\, [(q-1)\bar{N}+1,\ldots, q\bar{N}]} \in \bb{C}^{M_T \times \bar{N}}, \,\, q = 1,\ldots,Q\\ 
    \ma{G}^{(q)} &= \ma{G}_{.\, [(q-1)\bar{N}+1,\ldots, q\bar{N}]} \in \bb{C}^{M_R \times \bar{N}},\,\, q =1,\ldots,Q.
\end{align}
Hence, channel matrices can be seen as a concatenation of smaller submatrices such that $\ma{H}=[\ma{H}^{(1)},\ldots, \ma{H}^{(Q)}] \in \bb{C}^{M_T \times \bar{N}Q}$ and $\ma{G}=[\ma{G}^{(1)},\ldots, \ma{G}^{(Q)}] \in \bb{C}^{M_R \times \bar{N}Q}$. Additionally, $\ma{S}^{(q)}_k \in \bb{C}^{N \times N}$ is the corresponding scattering matrix used in the $k$-th training block, and $\ma{B}_k$ is the additive noise in the receiver modeled as a complex Gaussian random variable with zero mean and unitary variance.


The design of BD-RIS training matrices was first addressed in \cite{CE_BD_RIS}, which proposed an orthogonal construction for $\bar{\ma{S}} \in \bb{C}^{K \times \bar{N}^2Q}$ under the block-length condition $K \geq \bar{N}^2Q$. Although this design is effective for cascaded-channel acquisition, it does not provide separate estimates of the Tx--RIS and RIS--Rx channels. A subsequent work \cite{ClerckX_TSP_CE} introduced a modified training strategy by modeling the BD-RIS training structure as a third-order tensor, enabling individual channel estimation with reduced training overhead.

\textit{Motivation}: In BD-RIS-assisted systems, the training scattering tensor is intrinsically high-dimensional, and designing it independently at each block leads to a large number of control variables and substantial reconfiguration overhead. This motivates a structured low-dimensional representation that preserves the essential multilinear coupling while reducing implementation complexity. In this work, we adopt a low-rank PARAFAC parameterization of the BD-RIS training tensor to achieve this objective. This modeling choice plays a dual role in the paper: it yields a hardware-feasible training design with fewer parameters, and it also induces a tractable tensor structure in the received pilots that can be exploited for channel estimation. In the following, we derive the tensor signal formulation and then build PARAFAC-based estimators for recovering the individual channel factors.

\vspace{-0.3cm}
\section{PARAFAC-based channel estimation} \label{Sec:TB_CE_methods}
This section formulates the proposed channel estimation method and its physical implications for the \ac{BD-RIS} architecture. By exploiting the tensor structure of the received signal, we formulate a PARAFAC-based iterative block alternating least-squares method that decouple the estimation of $\ma{G}$ and $\ma{H}$. Then, we discuss solution-existence conditions and their implications for system design.

\vspace{-0.2cm}
\subsection{Tensor signal formulation}
Starting from the signal model in (\ref{eq:rec_sig_k_sum}), define $\ma{F}^{(q)} = \ma{X}\ma{H}^{(q)} \in \bb{C}^{T \times \bar{N}}$ for $q \in \{1,\ldots,Q\}$. By omitting the noise term for notational simplicity, the pilot signal received in the $k$-th block can be written as
\begin{equation}
    \label{eq:rec_sig_k_sum2} \ma{Y}_k = \sumx{q=1}{Q} \ma{G}^{(q)}\ma{S}^{(q)}_{k}\ma{F}^{(q)\text{T}} \in \bb{C}^{M_R \times T}.
\end{equation}
Equation (\ref{eq:rec_sig_k_sum2}) can be interpreted as the $k$-th frontal slice of a block Tucker-2 decomposition \cite{FavierAlmeida2014,andre_BTucker_2009} of the received tensor $\ten{Y} \in \bb{C}^{M_R \times T \times K}$. Therefore, using $n$-mode product notation, the received pilot signal tensor is expressed as
\begin{align}
    \label{eq:tenY}  \ten{Y}  = \sumx{q=1}{Q}  \ten{S}^{(q)}  \times_1 \ma{G}^{(q)} \times_2 \ma{F}^{(q)}, 
\end{align}
where $\ten{Y} = [\ma{Y}_{1} \sqcup_3 \ma{Y}_{2} \sqcup_3 \ldots \sqcup_3 \ma{Y}_{K} ] \in \bb{C}^{M_R\times T \times K}$ and $\ten{S}^{(q)} = [\ma{S}^{(q)}_{1} \sqcup_3 \ma{S}^{(q)}_{2} \sqcup_3 \ldots \sqcup_3 \ma{S}^{(q)}_{K} ] \in \bb{C}^{\bar{N} \times \bar{N} \times K}$. Each $\ten{S}^{(q)}$, $q=1,\ldots,Q$, is obtained by stacking the $q$-th BD-RIS scattering matrices along the third mode. We refer to
$\{\ten{S}^{(1)}, \ldots, \ten{S}^{(Q)}\}$ as the set of \emph{scattering tensors} associated with the $Q$ BD-RIS training groups. By combining these $Q$ tensor blocks, (\ref{eq:tenY}) can also be written as the following Tucker-2 decomposition:
\begin{equation}
\label{eq:tenY2}  \ten{Y} = \ten{S}  \times_1 \ma{G} \times_2 \ma{F} 
, 
\end{equation}
where $\ten{S}\doteq \big[\textrm{bdiag}\big(\ten{S}^{(1)}_{..1}, \ldots, \ten{S}^{(Q)}_{..1}\big) \sqcup_3 \ldots \sqcup_3  \textrm{bdiag}\big(\ten{S}^{(1)}_{..K}, \ldots, \ten{S}^{(Q)}_{..K}\big) \big]\in \mathbb{C}^{N \times N \times K}$ is a ``block-diagonal'' core tensor, each tensor block containing the BD-RIS scattering coefficients associated with a given group, and $\ma{F}=[\ma{F}^{(1)}, \ldots, \ma{F}^{(Q)}]=\ma{X}\ma{H} \in \mathbb{C}^{T \times \bar{N}Q}$. 

\subsection{PARAFAC decomposition of BD-RIS tensor}\label{Sec:Proposed_TB_CE}
We propose a structured PARAFAC design for the BD-RIS training tensor that keeps the interconnection topology fixed across blocks while allowing per-block adaptation through low-dimensional factors. We model each group tensor through the following third-order PARAFAC decomposition:
\begin{align}
\label{eq:tenS_PARAFAC}   \ten{S}^{(q)} = \ten{I}_{3,\bar{R}} \times_1 \bar{\ma{P}}_{1} \times_2 \bar{\ma{P}}_{2} \times_3 \ma{P}^{(q)}_{3} \in \bb{C}^{\bar{N} \times \bar{N} \times K},
\end{align}
$q = 1,\ldots,Q$, where $\bar{R}$ is the rank of the $q$-th group tensor, and $\bar{\ma{P}}_{1} \in \bb{C}^{\bar{N}\times \bar{R}}$, $\bar{\ma{P}}_{2} \in \bb{C}^{\bar{N}\times \bar{R}}$, and $\ma{P}^{(q)}_{3} \in \bb{C}^{\bar{K}\times \bar{R}}$ are its factor matrices. Unlike prior BD-RIS tensor designs that redesign the full scattering tensor at each block, this parameterization shares the spatial factors ($\bar{\ma{P}}_{1},\bar{\ma{P}}_{2}$) and updates only the block-dependent factor $\ma{P}^{(q)}_{3}$, which is the key mechanism behind the proposed complexity reduction.

Substituting (\ref{eq:tenS_PARAFAC}) into (\ref{eq:tenY}) and applying standard $n$-mode product manipulations, the noiseless received signal tensor is written as
\begin{align}
\notag \ten{Y} &= \sumx{q=1}{Q} \big( \ten{I}_{3,\bar{R}} \times_1 \bar{\ma{P}}_{1} \times_2 \bar{\ma{P}}_{2} \times_3 \ma{P}^{(q)}_{3}\big) \times_1 \ma{G}^{(q)} \times_2 \ma{F}^{(q)} \\
   \label{eq:tenY_PARAFAC} &= \sumx{q=1}{Q}  \ten{I}_{3,\bar{N}} \times_1 \ma{G}^{(q)}\bar{\ma{P}}_{1} \times_2 \ma{X}\ma{H}^{(q)}\bar{\ma{P}}_2 \times_3 \ma{P}^{(q)}_{3},
\end{align}
or, in compact form,
\begin{align}
    \label{eq:tenY_PARAFAC_total} \ten{Y} = \ten{I}_{3,\bar{N}Q} \times_1 \ma{G}\ma{P}_1\times_2 \ma{X}\ma{H}\ma{P}_2 \times_3 \ma{P}_{\text{S}} \in \bb{C}^{M_R \times T \times K},
 \end{align}
where 
\begin{align}
    \ma{P}_i =  \text{bdiag}\big(\bar{\ma{P}}_i,\ldots,\bar{\ma{P}}_i\big)  \in \bb{C}^{\bar{N}Q \times \bar{N}Q}, \; \text{for} \; i \in \{1,2\}
\end{align}
and
     $\ma{P}_{\text{S}} = [\ma{P}^{(1)}_{3},\ldots, \ma{P}^{(Q)}_{3}] \in \bb{C}^{K \times \bar{N}Q}$, 
\textcolor{black}{where $\bar{\ma{P}}_i$ follows a normalized truncated DFT design and $\ma{P}_S$ is designed as a random phase-mapping matrix with a uniform phase distribution.}
Equation (\ref{eq:tenY_PARAFAC_total}) makes the identifiability structure explicit: the unknown channels appear in separable multilinear factors, which directly enables decoupled estimation from tensor unfoldings.
From (\ref{eq:tenY_PARAFAC_total}), we derive its three unfoldings 
\begin{align}
    \label{eq:tenY_1_parafac}\nmode{Y}{1} &= \ma{G}\ma{P}_1\big( \ma{P}_{\text{S}} \diamond \ma{X}\ma{H}\ma{P}_2\big)^{\text{T}} \in \bb{C}^{M_R \times T K}, \\
    \label{eq:tenY_2_parafac}\nmode{Y}{2} &= \ma{X}\ma{H}\ma{P}_2\big( \ma{P}_{\text{S}} \diamond \ma{G}\ma{P}_1\big)^{\text{T}} \in \bb{C}^{T \times  M_R K}, \\
    \label{eq:tenY_3_parafac}\nmode{Y}{3} &= \ma{P}_{\text{S}}\big( \ma{X}\ma{H}\ma{P}_2 \diamond \ma{G}\ma{P}_1\big)^{\text{T}} \in \bb{C}^{K \times M_RT }.  
\end{align}




\subsection{PARAFAC alternating least squares (PALS) receiver}\label{sec:pals}
In this subsection, we develop a low-complexity PARAFAC alternating least-squares (PALS) receiver for decoupled estimation of the channel matrices. In the presence of noise, and based on (\ref{eq:tenY_PARAFAC_total}), the estimation problem is formulated as
\begin{equation}
\label{eq:Prob_PARAFAC}\underset{\ma{G},\ma{H}}{\text{min}}\Big\|\ten{Y} -  \ten{I}_{3,\bar{N}Q} \times_1 \ma{G}\ma{P}_1\times_2 \ma{X}\ma{H}\ma{P}_2 \times_3 \ma{P}_{\text{S}}
\Big\|_{\text{F}}^{2}.
\end{equation}
To solve (\ref{eq:Prob_PARAFAC}), we adopt the standard \ac{ALS} strategy \cite{comon2009tensor}. Specifically, we estimate $\ma{G}\in \bb{C}^{M_R \times \bar{N}Q}$ and $\ma{H} \in \bb{C}^{M_T \times \bar{N}Q}$ by alternating between the 1-mode and 2-mode unfoldings in (\ref{eq:tenY_1_parafac}) and (\ref{eq:tenY_2_parafac}), which leads to the following LS subproblems:
\begin{align*}
     \hat{\ma{G}}  &=  \underset{\ma{G}\ma{P}_1}{\argmin} \fronorm{\nmode{Y}{1} - \ma{G}\ma{P}_1\big( \ma{P}_{\text{S}} \diamond (\ma{X}\ma{H}\ma{P}_2)  \big)^{\text{T}}}^2 \\
     \hat{\ma{H}}  &=  \underset{\ma{H}}{\argmin} \fronorm{\nmode{Y}{2}- \ma{X}^{\text{H}}\ma{H}\ma{P}_2\big( \ma{P}_{\text{S}} \diamond \ma{G}\ma{P}_1\big)^{\text{T}}}^2,
\end{align*}
whose closed-form updates are
 \begin{align}
     \label{eq:sol_ZG}   \hat{\ma{G}} &= \nmode{Y}{1}\left[\ma{P}_1\big( \ma{P}_{\text{S}} \diamond (\ma{X}\ma{H}\ma{P}_2) \big)^{\text{T}} \right]^{+}, \\
          \label{eq:sol_ZF}   \hat{\ma{H}} &= \ma{X}^{\text{H}}\nmode{Y}{2}\left[\ma{P}_2\big( \ma{P}_{\text{S}} \diamond \ma{G}\ma{P}_1\big)^{\text{T}} \right]^{+}.
 \end{align}
The PALS receiver alternates between these two updates until convergence, similarly to BTALS \cite{ClerckX_TSP_CE}. At convergence, the estimates satisfy $\hat{\ma{G}} = \ma{G}\ma{\Lambda}_{\text{G}}$ and $\hat{\ma{H}} = \ma{H}\ma{\Lambda}_{\text{H}}$, where $\ma{\Lambda}_{\text{G}},\ma{\Lambda}_{\text{H}} \in \bb{C}^{N \times N}$ are diagonal scaling matrices such that $\ma{\Lambda}_{\text{G}}\ma{\Lambda}_{\text{H}}= \ma{I}_{N}$. These intrinsic scaling ambiguities do not affect BD-RIS optimization, which depends on the combined channel. The  procedure is summarized in Algorithm \ref{algorithm_bals_PARAFAC}.

 \begin{algorithm}[!t]
 \small
	\begin{algorithmic}[1]
		\setlength{\itemsep}{0pt}
		\caption{PARAFAC alternating least squares (PALS)}\label{algorithm_bals_PARAFAC}
		\State \textbf{Inputs}: received tensor $\ten{Y}$, BD-RIS factors $\ma{P}_{1}$, $\ma{P}_{2}$, $\ma{P}_{\text{S}}$, pilot matrix $\ma{X}$, maximum number of iterations $I$, and threshold $\eta$.
		\State Set $i=0$ and randomly initialize $\hat{\ma{H}}_{(0)}$.
\For{$i = 1:I$}		
		\State Update $\hat{\ma{G}}_{(i)}$ via LS as
          \vspace{-2ex}
  \begin{equation*}
\hat{\ma{G}}_{(i)} = \nmode{Y}{1}\left[\ma{P}_1\big( \ma{P}_{\text{S}} \diamond (\ma{X}\hat{\ma{H}}_{(i-1)}\ma{P}_2) \big)^{\text{T}} \right]^{+}
  \end{equation*}
     \vspace{-2ex}
	  \State Update $\hat{\ma{H}}_{(i)}$ via LS as
         \vspace{-2ex}
   \begin{equation*}
  \hat{\ma{H}}_{(i)}  = \ma{X}^{\text{H}}\nmode{Y}{2}\left[\ma{P}_2\big( \ma{P}_{\text{S}} \diamond \hat{\ma{G}}_{(i)}\ma{P}_1\big)^{\text{T}} \right]^{+}
   \end{equation*}
   \vspace{-2ex}
   \State Compute the reconstructed unfolding
   \begin{equation*}
   \hat{\ma{Y}}_{(i)}= \ma{P}_{\text{S}}\big((\ma{X}\hat{\ma{H}}_{(i)}\ma{P}_2)\diamond  (\hat{\ma{G}}_{(i)}\ma{P}_1) \big)^{\text{T}}
   \end{equation*}

   \State Compute the error $\epsilon_{(i)} = \Big\|\nmode{Y}{3}-\hat{\ma{Y}}_{(i)} \Big\|_{\text{F}}^{2}$.
 
    \State Check convergence and stop if $|\epsilon_{(i)} - \epsilon_{(i-1)}| \leq \eta$.
 \EndFor
		\State Return $\hat{\ma{G}}_{(i)} $ and $\hat{\ma{H}}_{(i)}$.
	\end{algorithmic}
\end{algorithm}



\subsection{Identifiability conditions and BD-RIS design complexity}

\begin{figure}
    \centering
    \includegraphics[scale = 0.6]{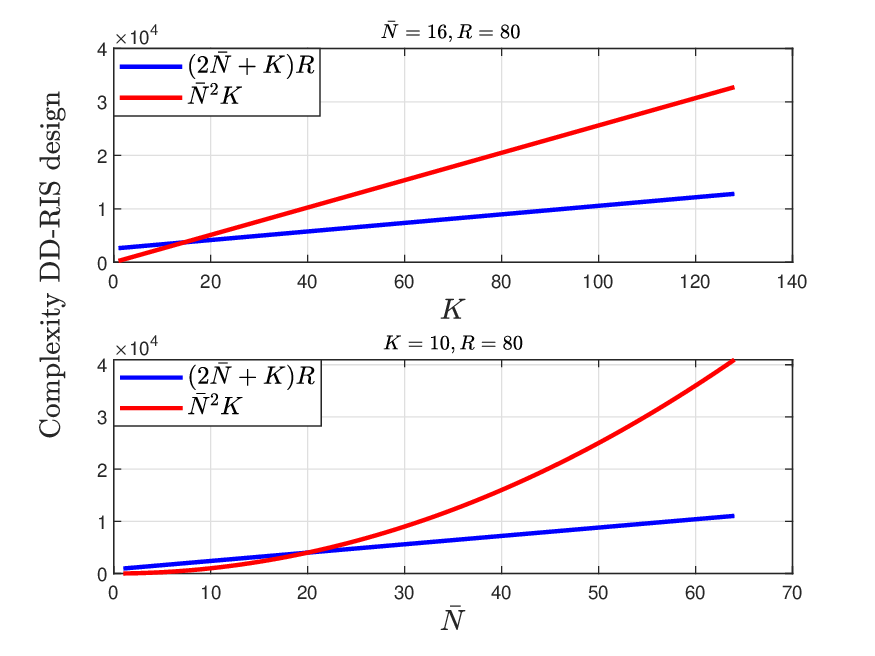}
    \caption{Comparison of BD-RIS training complexity: PARAFAC-based design vs. competing designs \cite{ClerckX_TSP_CE,sokal_asilomar}.}
    \label{fig:complexity}
\end{figure}

For the PALS receiver, uniqueness of the channel estimates obtained from (\ref{eq:sol_ZG}) and (\ref{eq:sol_ZF}) requires the following conditions:
\begin{itemize}
    \item From (\ref{eq:sol_ZG}),
    $\ma{M}_1 = \ma{P}_1\big( \ma{P}_{\text{S}} \diamond (\ma{X}\ma{H}\ma{P}_2)\big)^{\text{T}} \in \bb{C}^{\bar{N}Q \times KT}$
    must be full row-rank, implying $KT \geq \bar{N}Q$.
    \item From (\ref{eq:sol_ZF}),
    $\ma{M}_2 = \ma{P}_2\big(\ma{P}_{\text{S}} \diamond \ma{G}\ma{P}_1\big)^{\text{T}} \in \bb{C}^{\bar{N}Q \times KM_R}$
    must be full row-rank, and $\ma{X} \in \bb{C}^{T \times M_T}$ must be full column-rank. Therefore, $KM_R \geq \bar{N}Q$ and $T \geq M_T$.
\end{itemize}
Combining these conditions, a sufficient requirement on the number of blocks is
 $K \geq \max\big(\dfrac{\bar{N}Q}{T}, \dfrac{\bar{N}Q}{M_R}\big).
    \label{EQ: K_min}$
Since $T \geq M_T$ is always assumed, the minimum $K$ that guarantees identifiability is only $K \geq \bar{N}Q/M_R$. This is in contrast to LS (\cite{Li_TSP2025}) and BTKF (\cite{ClerckX_TSP_CE}) methods that require $K \geq \bar{N}^2Q$. Moreover, although the Tucker-based BTALS method of \cite{ClerckX_TSP_CE} can operate $K << \bar{N}^2Q$, a sufficient minimum condition on $K$ was not derived therein. We also emphasize that, unlike existing methods (\cite{Li_TSP2025,ClerckX_TSP_CE,sokal_asilomar,WangTSP2025}), which design the full BD-RIS tensor $\ten{S}^{(q)}$ directly, PALS operates on the factor matrices $\bar{\ma{P}}_{1} \in \bb{C}^{\bar{N} \times \bar{R}}$, $\bar{\ma{P}}_{2} \in \bb{C}^{\bar{N} \times \bar{R}}$, and $\{\ma{P}^{(q)}_{3}\} \in \bb{C}^{K \times \bar{R}}$, $q=1, \ldots, Q$. Consequently, the number of parameters to be adjusted is $(2\bar{N}+K)\bar{R}$. For instance, the methods in \cite{ClerckX_TSP_CE,sokal_asilomar} require adjusting $\bar{N}^2K$ scattering coefficients during training. Figure \ref{fig:complexity} compares the resulting design complexity with competing approaches. Note that, as the number of blocks increases, PALS becomes progressively more attractive in terms of complexity. A similar trend is observed when the number of elements per group increases, where the BD-RIS behavior approaches that of a diagonal (single-connected) RIS.

\vspace{-2ex}
\begin{remark}
    In this paper, we focus on the scenario where $R > N$; however, the proposed method is equally valid for the case $R \leq N$. In that situation, one can estimate the equivalent channels defined as $\overline{\ma{G}} = \ma{G}\ma{P}_1$ and $\overline{\ma{H}} = \ma{H}\ma{P}_2$. Nevertheless, the main objective of the paper is to recover the actual channels, i.e., without the masking matrices $\ma{P}_1$ and $\ma{P}_2$.
\end{remark}
\begin{remark}
 Using the 3-mode unfolding in (\ref{eq:tenY_3_parafac}), a closed-form solution can be used to recover the individual channels. This approach involves using a Khatri–Rao factorization to estimate the equivalent channels $\overline{\ma{G}}$ and $\overline{\ma{H}}$. Alternatively, a Kronecker factorization can be considered to directly estimate $\ma{G}$ and $\ma{H}$.
\end{remark}

\section{Numerical results}\label{Sec:Numerical_Results}
We assess the effectiveness of the proposed PALS receiver against representative baselines, namely LS, BTALS, and BTKF. The comparison is carried out in terms of \ac{NMSE} for the combined channel $\ma{C} = \ma{H} \blockron \ma{G}$ and for the individual factors. The \ac{NMSE} is defined as
$
    \text{NMSE}\big( \hat{\ma{\Omega}}_{\text{x}} \big) = \frac{\fronorm{\ma{\Omega} - \hat{\ma{\Omega}}}^2 }{\fronorm{\ma{\Omega}}^2},
$
where $\ma{\Omega} \in \{\ma{G}, \ma{H}, \ma{H}_C\}$. The BD-RIS tensor rank is defined as $R = \bar{R}Q$, i.e., the product between the rank of each group and the total number of groups.

\begin{figure}[t] 
     \centering
     \begin{subfigure}[b]{0.46\textwidth}
         \centering
         \includegraphics[width=\textwidth]{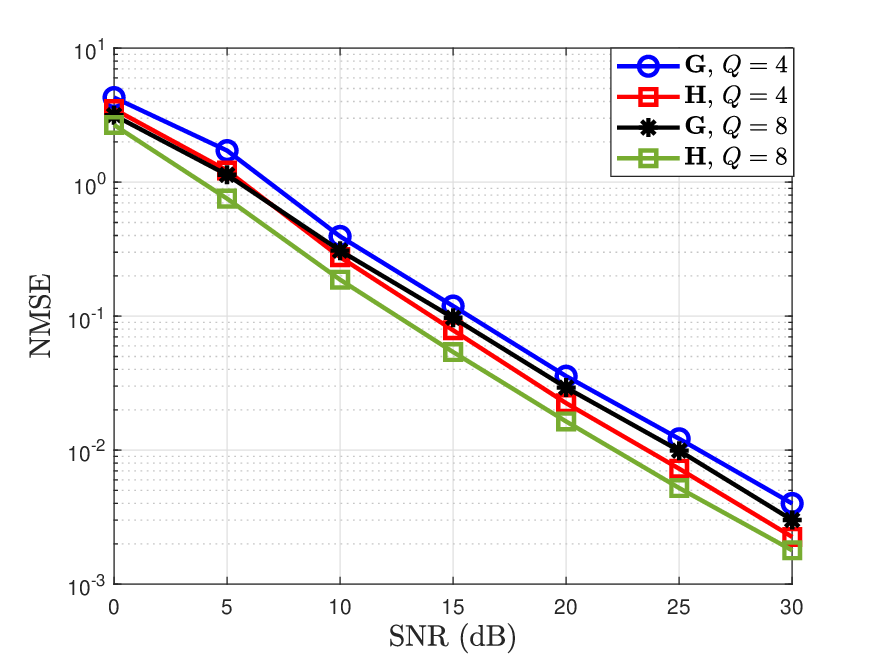}
         \caption{NMSE of individual channels.}
         \label{fig:NMSE}
     \end{subfigure}
     \hfill 
   \begin{subfigure}[b]{0.46\textwidth}
         \centering
         \includegraphics[width=\textwidth]{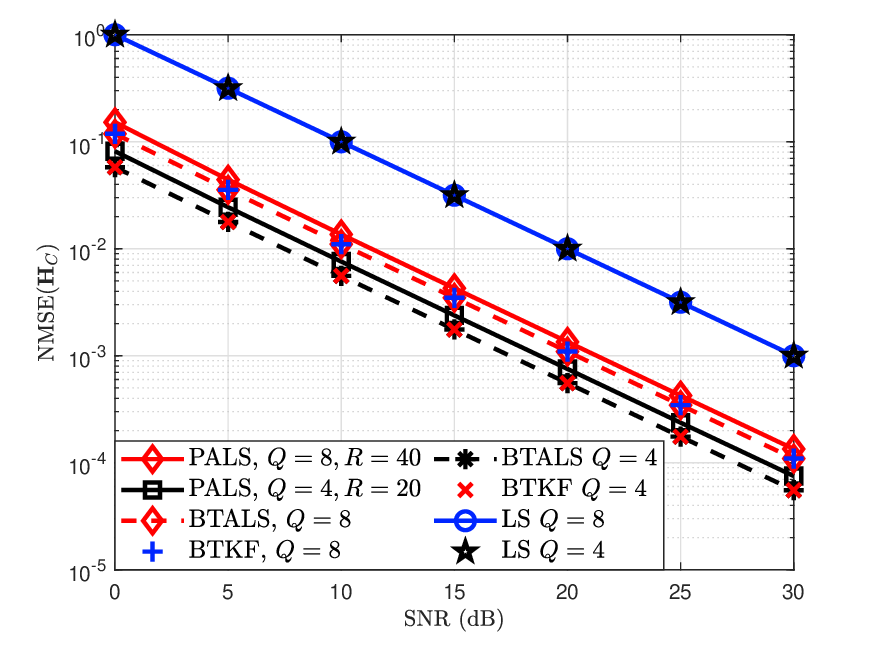}
         \caption{Comparison with state-of-the-art.}
         \label{fig:HC}
     \end{subfigure}
    
     \caption{Experimental results: (a) NMSE of $\mathbf{G}$ and $\mathbf{H}$, (b) Comparison with competing methods.}
     \label{fig:geral_comparativo}
\end{figure}

Figure \ref{fig:NMSE} reports the estimation accuracy of PALS for the individual channels $\ma{G}$ and $\ma{H}$ under $N = 16$, $M_r = 10$, $M_t = T = 6$, $K = 10$, and $\bar{R} = 5$. Two important trends emerge. First, PALS yields accurate recovery of both factors, with slightly better estimation for $\ma{H}$. Second, performance improves as the number of groups increases, which is consistent with the rank increase $R=\bar{R}Q$ and the associated diversity gain. These results indicate that the proposed low-rank training design does not compromise estimation fidelity. Figure \ref{fig:HC} compares PALS with prior channel-estimation strategies. The gain over conventional LS \cite{ClerckX_TSP_CE} is clear, confirming that explicitly exploiting multilinear structure is essential in BD-RIS training. More importantly, PALS achieves accuracy comparable to BTALS/BTKF \cite{ClerckX_TSP_CE,sokal_asilomar}, which are strong tensor baselines. This is a central result of the paper: PALS preserves state-of-the-art estimation performance while relying on a substantially more structured and implementation-oriented BD-RIS design. 

Beyond NMSE, PALS offers a practical advantage. In BTALS/BTKF, the BD-RIS tensor is redesigned at every block $k$, implying repeated reconfiguration of inter-element connections. In contrast, PALS keeps the interconnection topology fixed and updates only phase coefficients. This design reduces control complexity (Figure \ref{fig:complexity}) and makes hardware realization significantly more feasible. 
In summary, these results demonstrate that PALS is not only accurate, but also markedly more scalable and implementation-ready compared with competing methods.



\section{Conclusions}
The central novelty of the PARAFAC-based channel estimation framework is a structured low-rank design of the BD-RIS training tensor in which the interconnection topology is kept fixed across training blocks and only low-dimensional phase-related factors are updated. This design departs from prior tensor-based BD-RIS methods that rely on repeated full-tensor reconfiguration and therefore incur higher implementation complexity. Building on this design, we developed the PALS receiver to estimate the individual channels from the induced multilinear model. Additionally, we derived the minimum number of training blocks, a result not previously established in the literature. Our numerical results show that PALS substantially outperforms conventional LS, matches the estimation accuracy of state-of-the-art tensor receivers (BTALS/BTKF), and simultaneously reduces BD-RIS design and control complexity. Hence, PALS provides an effective accuracy--complexity tradeoff, making high-performance BD-RIS channel estimation more practical for implementation. Future work includes extending the framework to hardware-impairment-aware models.
\bibliographystyle{IEEEtran}
\bibliography{ref.bib}

@ARTICLE{ClerckX_TSP_CE,
  author={de Almeida, André L. F. and Sokal, Bruno and Li, Hongyu and Clerckx, Bruno},
  journal={IEEE Transactions on Signal Processing}, 
  title={Channel Estimation for Beyond Diagonal RIS via Tensor Decomposition}, 
  year={2025},
  volume={73},
  number={},
  pages={4764-4779},
  }

@article{FavierAlmeida2014,
author = {Favier, G. and de Almeida, A. L. F.},
title = {Overview of constrained {PARAFAC} models},
journal = {EURASIP Journal of Advances in Signal Processing},
volume = {2014},
number = {142},
pages = {1-25},
month={Sep.},
year = {2014},
}

@Article{Harshman,
  author  = {R. A. Harshman},
  title   = {Foundations of the {PARAFAC} procedure: Models and conditions for an ``explanatory'' multi-modal factor analysis},
  journal = {UCLA Working Papers in Phonetics},
  year    = {1970},
  volume  = {16},
  pages   = {1-84},
}

@article{gil2021,
  author={de Araújo, Gilderlan T. and de Almeida, André L. F. and Boyer, Rémy},
  journal={IEEE J. Sel. Topics Signal Process.}, 
  title={Channel Estimation for Intelligent Reflecting Surface Assisted {MIMO} Systems: A Tensor Modeling Approach}, 
  year={2021},
  month={Feb.},
  volume={15},
  number={3},
  pages={789-802},
  doi={10.1109/JSTSP.2021.3061274}
  }

@article{comon2009tensor,
  title={{Tensor decompositions, alternating least squares and other tales}},
  author={Comon, Pierre and Luciani, Xavier and de Almeida, Andr{\'e} L. F.},
  journal={Journal of Chemometrics},
  volume={23},
  number={7-8},
  pages={393--405},
  year={2009},
  publisher={Wiley Online Library}
}

@INPROCEEDINGS{CE_BD_RIS,
  author={Li, Hongyu and Zhang, Yumeng and Clerckx, Bruno},
  booktitle={(CAMSAP)},
  title={{Channel estimation for beyond diagonal reconfigurable itntelligent surfaces with group-connected architectures}}, 
  year={2023},
  volume={},
  number={},
  pages={21-25}}

@article{andre_BTucker_2009,
  title={{Constrained Tucker-3 model for blind beamforming}},
  author={de Almeida, Andre L. F. and Favier, G{\'e}rard and Mota, Jo{\~a}o C. M},
  journal={Signal Processing},
  volume={89},
  number={6},
  pages={1240--1244},
  year={2009},
  publisher={Elsevier}
}

@ARTICLE{CLX_BD_GC_2023,
  author={Li, Hongyu and Shen, Shanpu and Clerckx, Bruno},
  journal={IEEE Trans. Wireless Commun.}, 
  title={Beyond Diagonal Reconfigurable Intelligent Surfaces: From Transmitting and Reflecting Modes to Single-, Group-, and Fully-Connected Architectures}, 
  year={2023},
  volume={22},
  number={4},
  pages={2311-2324},
  doi={10.1109/TWC.2022.3210706}}

@ARTICLE{CLX_MS_2023,
  author={Li, Hongyu and Shen, Shanpu and Clerckx, Bruno},
  journal={IEEE J. Sel. Areas Commun.}, 
  title={Beyond Diagonal Reconfigurable Intelligent Surfaces: A Multi-Sector Mode Enabling Highly Directional Full-Space Wireless Coverage}, 
  year={2023},
  volume={41},
  number={8},
  pages={2446-2460},
  doi={10.1109/JSAC.2023.3288251}}

@INPROCEEDINGS{sokal_asilomar,
  author={Sokal, Bruno and Fazal-E-Asim and de Almeida, André L. F. and Li, Hongyu and Clerckx, Bruno},
  booktitle={(ASILOMAR)}, 
  title={A Decoupled Channel Estimation Method for Beyond Diagonal {RIS}}, 
  year={2024},
  volume={},
  number={},
  pages={1395-1399},
  doi={10.1109/IEEECONF60004.2024.10942770}}

@INPROCEEDINGS{gil-asilomar,
  author={de Araújo, Gilderlan Tavares and de Almeida, André L. F.},
  booktitle={(ASILOMAR)}, 
  title={Semi-Blind Channel Estimation for Beyond Diagonal RIS}, 
  year={2024},
  volume={},
  number={},
  pages={1586-1590},
}

@misc{gil_OJSP,
      title={Semi-Blind Joint Channel and Symbol Estimation for Beyond Diagonal Reconfigurable Surfaces}, 
      author={Gilderlan Tavares de Araújo and André L. F. de Almeida and Buno Sokal and Gabor Fodor and Paulo R. B. Gomes},
      year={2025},
      eprint={2512.15441},
      archivePrefix={arXiv},
      primaryClass={eess.SP},
      url={https://arxiv.org/abs/2512.15441}, 
}

@INPROCEEDINGS{Araujo_SAM20,
  author={de Araújo, Gilderlan T. and de Almeida, André L. F.},
  booktitle={2020 IEEE 11th Sensor Array and Multichannel Signal Processing Workshop (SAM)}, 
  title={PARAFAC-Based Channel Estimation for Intelligent Reflective Surface Assisted MIMO System}, 
  year={2020},
  volume={},
  number={},
  pages={1-5},
  doi={10.1109/SAM48682.2020.9104260}}

@ARTICLE{Paulo_2023,
  author={Gomes, Paulo R. B. and de Araújo, Gilderlan Tavares and Sokal, Bruno and Almeida, André L. F. de and Makki, Behrooz and Fodor, Gábor},
  journal={IEEE Transactions on Vehicular Technology}, 
  title={Channel Estimation in {RIS}-Assisted {MIMO} Systems Operating Under Imperfections}, 
  year={2023},
  volume={72},
  number={11},
  pages={14200-14213},
  doi={10.1109/TVT.2023.3279805}}

@ARTICLE{Li_TSP2025,
  author={Li, Hongyu and Shen, Shanpu and Zhang, Yumeng and Clerckx, Bruno},
  journal={IEEE Transactions on Signal Processing}, 
  title={Channel Estimation and Beamforming for Beyond Diagonal Reconfigurable Intelligent Surfaces}, 
  year={2024},
  volume={72},
  number={},
  pages={3318-3332},
  doi={10.1109/TSP.2024.3424229}}

@ARTICLE{WangTSP2025,
  author={Wang, Rui and Zhang, Shuowen and Clerckx, Bruno and Liu, Liang},
  journal={IEEE Transactions on Signal Processing}, 
  title={Low-Overhead Channel Estimation Framework for Beyond Diagonal Reconfigurable Intelligent Surface Assisted Multi-User MIMO Communication}, 
  year={2025},
  volume={73},
  number={},
  pages={4700-4717},
  doi={10.1109/TSP.2025.3628337}}

@ARTICLE{Lee_2022,
  author={Swindlehurst, A. Lee and Zhou, Gui and Liu, Rang and Pan, Cunhua and Li, Ming},
  journal={Proceedings of the IEEE}, 
  title={Channel Estimation With Reconfigurable Intelligent Surfaces—A General Framework}, 
  year={2022},
  volume={110},
  number={9},
  pages={1312-1338},
  doi={10.1109/JPROC.2022.3170358}}

@ARTICLE{Di_Renzo2020,
  author={Di Renzo, Marco and Zappone, Alessio and Debbah, Merouane and Alouini, Mohamed-Slim and Yuen, Chau and de Rosny, Julien and Tretyakov, Sergei},
  journal={IEEE J. Sel. Areas Commun.}, 
  title={Smart Radio Environments Empowered by Reconfigurable Intelligent Surfaces: How It Works, State of Research, and The Road Ahead}, 
  year={2020},
  volume={38},
  number={11},
  pages={2450-2525},
  doi={10.1109/JSAC.2020.3007211}}

@ARTICLE{Rui_Zhang_2021,
  author={Wu, Qingqing and Zhang, Shuowen and Zheng, Beixiong and You, Changsheng and Zhang, Rui},
  journal={IEEE Trans. Commun.}, 
  title={Intelligent Reflecting Surface-Aided Wireless Communications: A Tutorial}, 
  year={2021},
  volume={69},
  number={5},
  pages={3313-3351},
  doi={10.1109/TCOMM.2021.3051897}}

\end{document}